\documentclass[conference]{IEEEtran}
\IEEEoverridecommandlockouts
\usepackage{cite}
\usepackage{amsmath,amssymb,amsfonts}
\usepackage{algorithmic}
\usepackage{graphicx}
\usepackage{textcomp}
\usepackage{xcolor}
\def\BibTeX{{\rm B\kern-.05em{\sc i\kern-.025em b}\kern-.08em
    T\kern-.1667em\lower.7ex\hbox{E}\kern-.125emX}}
    
\usepackage{multirow}
\usepackage{hyperref}
\usepackage{setspace}

\makeatletter
\def\@IEEEpubidpullup{8\baselineskip}
\makeatother

\usepackage{fancyhdr}
\usepackage{kantlipsum}
\fancypagestyle{plain}{
\fancyhf{}
\fancyhead[C]{Conference on \LaTeX} 

}
\usepackage{eso-pic}

\begin{document}

\IEEEoverridecommandlockouts
\IEEEpubid{
\parbox{\columnwidth}{\vspace{-4\baselineskip} Permission to make digital or hard copies of all or part of this work for personal or classroom use is granted without fee provided that copies are not made or distributed for profit or commercial advantage and that copies bear this notice and the full citation on the first page. Copyrights for components of this work owned by others than ACM must be honored. Abstracting with credit is permitted. To copy otherwise, or republish, to post on servers or to redistribute to lists, requires prior specific permission and/or a fee. Request permissions from \href{mailto:permissions@acm.org}{permissions@acm.org}.\hfill\vspace{-0.8\baselineskip}\\
\begin{spacing}{1.2}
\small\textit{ASONAM '23}, November 6-9, 2023, Kusadasi, Turkey \\
\copyright\space2023 Association for Computing Machinery. \\
ACM ISBN 979-8-4007-0409-3/23/11\ldots\$15.00 \\
\url{https://doi.org/10.1145/3625007.3627512}
\end{spacing}
\hfill}
\hspace{0.9\columnsep}\makebox[\columnwidth]{\hfill}}
\IEEEpubidadjcol
\bstctlcite{MyBSTcontrol}

\title{PureNav: A Personalized Navigation Service for Environmental Justice Communities Impacted by Planned Disruptions}



\author{\IEEEauthorblockN{1\textsuperscript{st} Omar Hammad}
\IEEEauthorblockA{\textit{Information and Computer Science Dept.} \\
\textit{King Fahd University of Petroleum and Minerals}\\
Dhahran, Saudi Arabia \\
omarjh@kfupm.edu.sa}
\and
\IEEEauthorblockN{2\textsuperscript{nd} Md Rezwanur Rahman}
\IEEEauthorblockA{\textit{Computer Science Dept.} \\
\textit{University of Colorado, Boulder}\\
Boulder, USA \\
mdra7255@colorado.edu }
\and
\IEEEauthorblockN{3\textsuperscript{rd} Nicholas Clements}
\IEEEauthorblockA{\textit{Mechanical Engineering Dept.} \\
\textit{University of Colorado, Boulder}\\
Boulder, USA \\
nicholas.clements@colorado.edu}
\and
\IEEEauthorblockN{4\textsuperscript{th} Shivakant Mishra}
\IEEEauthorblockA{\textit{Computer Science Dept.} \\
\textit{University of Colorado, Boulder}\\
Boulder, USA \\
mishras@cs.colorado.edu}
\and
\IEEEauthorblockN{5\textsuperscript{th} Shelly Miller}
\IEEEauthorblockA{\textit{Mechanical Engineering Dept.} \\
\textit{University of Colorado, Boulder}\\
Boulder, USA \\
shelly.miller@colorado.edu}
\and
\IEEEauthorblockN{6\textsuperscript{th} Esther Sullivan}
\IEEEauthorblockA{\textit{Sociology Dept.} \\
\textit{University of Colorado, Denver}\\
Denver, USA \\
esther.sullivan@ucdenver.edu}
}

\maketitle

\begin{abstract}
Planned disruptions such as highway constructions are commonplace nowadays and the communities living near these disruptions generally tend to be environmental justice communities---low socio-economic status with disproportionately high and adverse human health
and environmental effects. A major concern is that such activities negatively impact people's well-being by disrupting their daily commutes via frequent road closures and increased dust \& air pollution. This paper addresses this concern by developing a personalized navigation service called PureNav to mitigate the negative impacts of disruptions in daily commutes on people’s well-being. PureNav has been designed using active engagement with four environmental justice communities affected by major highway construction. It has been deployed in the real world among the members of the four communities, and a detailed analysis of the data collected from this deployment as well as surveys show that PureNav is potentially useful in improving people’s wellbeing. The paper describes the design, implementation, and evaluation of PureNav, and offers suggestions for further improving its efficacy.
\end{abstract}






\section{intoduction}

\par The Central 70 (C70) project in Northwest Denver, Colorado, USA was a major highway construction project that occurred over a period of about five years (2017-2022)\cite{cdot_central_2019}. Four communities (Elyriya-Swansea, Globeville, Cole, and Calyton) living in the vicinity of this construction site are  Environmental Justice Communities (EJC)---low socio-economic status with disproportionately high and adverse human health
and environmental effects\cite{bullard_environmental_1993}. The overall goal of our Social Justice and Environmental Quality - Denver (SJEQ) project ({\it https://www.sjeqdenver.com/}) is to first understand the negative impact of C70 on the health and well-being of these communities and then develop interventions to mitigate the negative impacts.

\par To learn firsthand the community’s local knowledge
and concerns regarding the C70 project, we organized three focus groups in the Summer of 2021 with the residents of the
four communities to discuss the topics of C70 construction and community concerns. Overall, 32 residents from these communities participated in our focus groups which included both English
and Spanish speakers. We obtained approval from IRB before conducting these focus groups. In addition, we have developed a smartphone app called PurEmotion to
understand the impact of the C70 construction project on
people’s well-being (See https://www.sjeqdenver.com/ and Section \ref{sec:study} for more information on PurEmotion).

Based on the results from the focus groups and two separate deployments of PurEmotion, each over a six-week duration among about 85-95 community participants, one major issue we identified that impacted negatively people's well-being is disruptions and inconvenience with their daily commutes. There were major concerns related to traffic and road closures due to construction,
including increased time spent in traffic, longer commutes, and delays. In addition, community members faced
consistent noise and increased pollution, including dirt and dust. Because of this, community members have been adjusting their schedules by allocating more time for traffic and detours, waking up kids earlier and leaving home earlier, and/or staying home
more. Many also changed their habits, including avoiding certain areas or going to different stores, keeping
windows closed, and spending less time outdoors.

\par To address these concerns, we have designed and developed a personalized navigation service called {\it PureNav} to mitigate the negative impacts of disruptions in daily commutes on people's well-being. In this paper, we describe the design and implementation of PureNav, our experience in deploying this service, and evaluation results from this deployment. We aim to answer the following research questions: 

\begin{enumerate}

    \item How do we build a helpful software system to support environmental justice communities while navigating the areas affected by large construction projects?

    \item What can we learn about Trip scheduling behavior from the usage of an information system deployed in environmental justice communities? 

    \item To what extent does the proposed intervention mechanism, PureConnect help in mitigating the negative impacts of the construction project in environmental justice communities who live in the affected areas?
    
\end{enumerate}

We developed the first version of PureNav as a smartphone app that recommended commute routes, provided step-by-step directions, and provided some additional personalized information about the commute route. Twenty-three community members used this smartphone app along with the PurEmotion app for six weeks. Feedback from the app participants was mixed. While the participants liked the personalized information, they found the step-by-step directions of the app not useful and distracting. The general feedback was that they didn't need these directions for the commutes that they undertake regularly. Based on this feedback, we have developed the second version of PureNav implemented as a slack workspace ({\it https://www.slack.com}). This version provides metadata related to different possible routes and personalized information about the routes in advance but does not provide any specific route recommendations or step-by-step directions. We deployed this new version among 41 community members who used the app for six weeks along with the PurEmotion app. Feedback from the participants after using the app is positive and evaluation from the data collected from this deployment (PureNav and PurEmotion) shows that there was an improvement in the wellbeing of people when they used PureNav. In particular, this paper makes the following important contributions:

\begin{itemize}
    \item PureNav is the first navigation service to the best of our knowledge, specifically designed to address the needs of environmental justice communities facing disruptions due to major construction activities in their vicinity.
    \item PureNav has been built and experimented with using active engagement with the environmental justice communities affected by a major construction.
    \item Multiple investigations in terms of real-world deployment and surveys show that environmental justice community users appreciate PureNav functionalities and this service is potentially useful in improving their well-being.
\end{itemize}


\section{Literature Review}

\par The most familiar smartphone navigation system is Google Maps \cite{anderson_more_2016} which has been chosen as the top navigation app since it introduced turn-by-turn navigation in 2008 \cite{samson_exploring_2019}. Many drivers have considered it, along with the Waze app, as an alternative to the in-car navigation system \cite{samson_exploring_2019}. At the same time, Waze has reported a very high monthly usage (65M people) in more than 185 countries \cite{samson_exploring_2019}. There are also some other popular apps like WeGo, HERE, MapQuest, and Bing Maps \cite{samson_exploring_2019}. 

\par Since sensors in smartphones have become much better, most of these apps use data from online app users to calculate the routing time and suggest the best routes for users \cite{samson_exploring_2019}. However, unlike traditional navigation methods, Waze has used crowdsourcing to report road problems  \cite{samson_exploring_2019} and introduced the idea of navigation among drivers and navigators as a social activity. The type of information that these apps provide falls into three categories, experiential, descriptive, and perspective\cite{ben-elia_response_2015} where most apps provide the latter two methods \cite{sha_social_2013}. 

\par Although existing navigation systems are useful in many scenarios, a more personalized system is needed for our project. For instance, existing apps focus primarily on recommending the fastest route although it was shown that this is not always the case for users \cite{samson_exploring_2019,zhu_people_2015}. Google Maps for example suggest the top fastest routes and lets users choose one of them. Moreover, Waze starts the fastest route without asking the users \cite{levine_system_2014}. A previous study showed how existing routing services are not aligned with local drivers's experience \cite{ceikute_routing_2013}. On top of that Waze and Google Maps don't offer the eco-friendly routes option \cite{samson_exploring_2019}. In a  study by Pfleging et. al \cite{pfleging_experience_2014} it was shown that among different routing preferences, almost half of the people preferred the fuel-efficient route, and only 18\% and 3.5\% preferred the fastest and shortest routes respectively. In addition, a couple of studies \cite{fujino_detecting_2018,brown_normal_2012} showed that a lot of people deviate from the recommended optimal route because of many reasons such as complex routes, inaccuracies in GPS \cite{fujino_detecting_2018}, and unfamiliarity of the road, especially if the time is negligible \cite{samson_exploring_2019}. In a study by Samson and Sumi, they reported that participants would only use the fastest route if it was 10 minutes faster. Further, Patel et al \cite{patel_personalizing_2006} showed that people prefer simple instructions for routing using familiar landmarks. Other mentioned routing preferences were less traffic, a straightforward path, and shorter distance. However, when people are in a hurry they choose the fastest route \cite{samson_exploring_2019}.

\par Existing systems were shown to be distracting \cite{boudette_biggest_2016,christensen_towards_2019}, and hard to use by older adults \cite{mcwilliams_effects_2015,yu_maps_2020}, which shapes a big part of our participants. Old adults tend to have lower visual and motor abilities due to the aging process \cite{carlson_aging_1995}. Despite having more driving experience, old adults have difficulty controlling the vehicle and using the phone at the same time. In addition, due to the high amount of instructions and reliability of in-car navigation systems, they tend not to follow the routing instructions all the time \cite{al_mahmud_user_2009}. Other factors that make these systems hard to use exist such as inadequate visual saliency \cite{yu_maps_2020} like font, ambiguous UI meaning, and low information scent.


\section{Study Design}
\label{sec:study}

\subsection{Pre and Post Intervention Phases}

\par The SJEQ-Denver project is structured over pre-intervention and post-intervention phases.The goal of pre the first is to understand the impact of C70 construction on people's well-being, while in the second phase, the goal is to introduce appropriate interventions and assess whether they result in improving people's well-being. The pre-intervention phase consisted of deploying the PurEmotion app over two different periods (called Cohort 1 and Cohort 2) to collect well-being, location, and air quality information from participants. PurEmotion is a smartphone diary app that community members use once a day to answer a few survey questions about their current feelings, their perceptions about the air quality around them, and recent experiences about their daily commutes. 

\par Based on our findings from Cohort 1 and 2, we introduced three interventions: (1) PureConnect service to increase awareness about the construction project, (2) indoor air cleaners to improve the indoor air quality, and (3) PureNav service to help community members navigate easier. These interventions were deployed over six weeks each (called Cohort 3 and Cohort 4). PurEmotion was again deployed in these cohorts to assess the effectiveness of these interventions.

While this paper focuses on PureNav, information
about PureConnect and indoor air cleaners is available at
{\it https://www.sjeqdenver.com/}.

\subsection{PureNav}

\subsubsection{System Design}

\par 
Although community members use Google Maps, Apple Maps, and other navigation apps, these apps do not provide instant information about road closures. In addition, these apps cause distractions for older people while driving \cite{boudette_biggest_2016,ceikute_routing_2013}, thus making them hard to be used by them \cite{lopez_behind_2017,sha_social_2013}. Moreover, given that C70 was causing environmental pollution and disruptions in a relatively small local area, we need a service that considers environmental factors as well as incorporates disruptions as they occur, and provides personalized information to each user based on their preference.
Based on these observations, we incorporated the following features in PureNav:


\begin{itemize}
    \item Authentication: Allow users to sign, log in, and log out from the system using their phone numbers
    \item Common place: Allow users to create, view, update, and delete places that they visit regularly. 
    \item Routing preferences: Allow users to choose their routing preferences (e.g. medium, fastest, or less polluted route).
    \item Route recommendation: Suggest the best route (based on preferences) between common places.
    \item Trips: Calculate how long it will take to go from locations A to B based on when the user needs to be at B and alert the user a few minutes earlier, the best time to start to reach B at the right time.
    \item Trip monitor: Monitor any change in traffic and alert a person if it changes.
    \item Rate trip: Allow users to rate a trip after they have completed the trip.
\end{itemize}

\subsubsection{PureNav (Version 1)}

\par The first version of PureNav was built using React Native ({\it https://reactnative.dev/}) as a front-end technology for both Android and iOS users.
The app starts by asking users to type in their phone numbers that are validated in the backend.
Next, users add common trips that they undertake, such as trips to the office, grocery store, etc.
For each trip, users provide attributes such as name, address, transportation medium (bus, car, etc.), days of trips (Monday, Tuesday, etc.), and arrival time. The app incorporates all best practices, such as a map view for addresses, autocompletion, and suggestions for common trip names like work, school, etc. Then the user lands on the home screen, which displays the next trip with a  \textit{Go Now} button and a list of future trips. On clicking this button, the user sees two options for the route: the fastest and the safest route. The {\it safest route} is the route with the best (average) air quality. 
After arriving at the destination, users are presented with a feedback screen that asks users to rate the overall experience of the trip from 1 to 5, what was good about the trip, what was bad about the trip, road closures along the road, and finally any text feedback that they want to share.




\subsubsection{Route and Time Calculations}



\par To calculate the fastest route we used Google Maps Directions API to return the suggested routes. However, if a road closure is reported, we designed an algorithm to suggest the fastest route that avoids reported road closures with the assistance of Google Maps Directions API using the \textit{waypoints} attribute. The algorithm takes source, destination, road closures, and medium into account. If the medium is not driving or there are no reported road closures, we use the fastest route from Google. Otherwise, we use the following algorithm:

\begin{itemize}
    \item Define a set of n points (100 by default) centered around the source point in a spiral shape 
    \item For each waypoint: 
    \begin{itemize}
        \item gets the fastest route from source to destination passing by the waypoint. 
        \item if the route does not cross the road closures, add it to a list of valid routes
    \end{itemize}
    \item calculates the duration for each route and returns the fastest.
\end{itemize}

\par To calculate the \textit{safest route} we used Breezometer (https://breezometer.com/) and Google Maps Directions APIs. The algorithm takes a list of routes, and for each route, it selects five points (latitude/longitude) along the route equally distant from one another. For all five points, we get PM2.5 concentration using the Breezometer API. The safest route recommended is the one with the lowest average PM2.5 concentration.


\par To calculate the \textit{time to start} we have designed an algorithm that takes source, destination, and preferred arrival time and uses Google Maps Directions APIs to calculate the duration under three scenarios: optimistic for light traffic, best for average traffic, and pessimistic for worst traffic. Based on this duration and preferred arrival time, we calculate the time to start based on current traffic if the arrival time is soon or pessimistic if the arrival time is further in the future. The pessimistic start time is updated as the time to start is closer.

\subsubsection{PureNav (Version 2)}

\par Based on the feedback we received after deploying version 1 in Cohort 3 among 21 users, we decided to focus on personalized information that would help users before they head to their destination instead of route recommendations. 
\par We migrated the front-end code to the slack workspace (see Fig. \ref{fig:purenav_v2}) with some changes to the app flow. In particular, the service provides the following information:

\begin{figure*}
    \centering
    \includegraphics[width=0.9\textwidth]{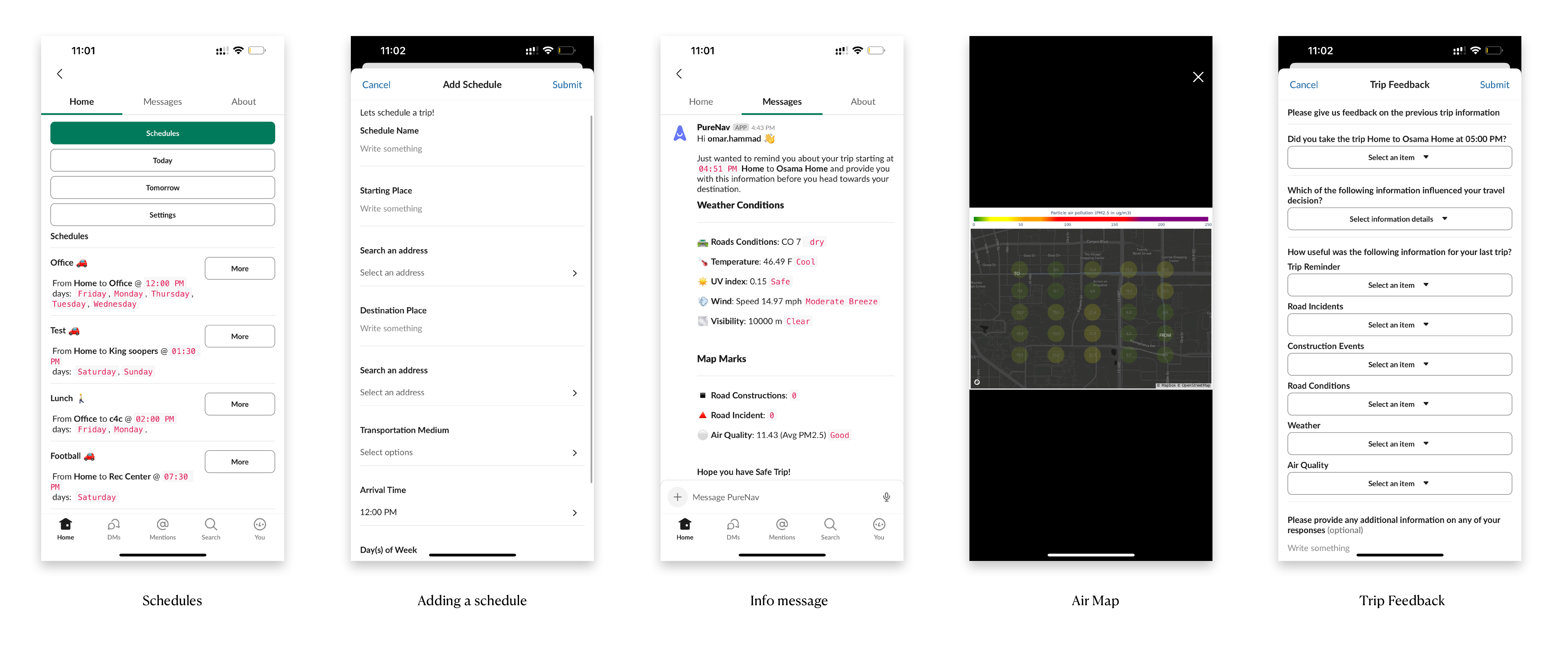}
    \caption{Main screen of the second version of PureNav after we received feedback from the 3rd cohort of the study. The figures highlight the list of schedules, the process of adding a schedule, trip reminders with related information, the air quality map for the road, and the feedback form.}
    \label{fig:purenav_v2}
\end{figure*}


\begin{itemize}
    \item \textbf{Trip Reminder}: Sends a reminder message on Slack before a trip is supposed to start (10, 15, or 30 minutes before the trip). The reminder includes Road Conditions, Temperature, UV Index, Wind, Visibility, Road Construction/Incidents, and Air Quality information. The reminder also includes a customized map image with PM 2.5 concentration. 
    \item \textbf{Road Incidents}: Road incidents information, e.g. accidents, is fetched from COTrip API ({\it https://www.cotrip.org/}) in real-time.
    \item \textbf{Construction Events}: Construction activities information is retrieved from the Colorado Department of Transportation ({\it https://www.codot.gov/}).
    \item \textbf{Road Conditions}: Road conditions are indicated in user-understandable tags e.g. snowy, slippery, dry.
    \item \textbf{Weather}: We list temperature with tags, Cold, Very Cold, Warm, Hot; UV index numbers with a precautionary message; Wind speed and direction;  Visibility
    \item \textbf{ Air Quality}: It is indicated as PM2.5 concentration.
\end{itemize}

\par The flow of the app changed slightly. First, we did not require users to sign up since the app is integrated into Slack. Second, for each trip, the app asks users to enter the source and destination information instead of just providing the destination information. Unlike the previous version which calculated trip duration and route dynamically with GPS, the second version does not have a way to track users' GPS location.

\par When the trip's departure time gets close the app sends a reminder to the user, but it does not recommend routes. Instead, it shows the user the list of information mentioned above and a map that previews the area between the source and destination along with the air quality (PM2.5 concentration) on different areas of the map.
Fifteen minutes after the projected arrival time, the app prompts the user to fill in a feedback form:

\begin{itemize}
    \item Did you take the trip A to B?
    \item Which of the following information influenced your travel decision? (Road incidents, construction events, road conditions, weather, air quality)
    \item How useful was the following information for your last trip (incidents, construction events, road conditions, weather, air quality)? (1: Not useful to 5: Very useful). 
    \item Please provide any additional information on any of your responses(optional).
\end{itemize}


\section{Evaluation}

PureNav Version 1 was deployed among 23 users in Cohort 3 (Oct-Dec 2022) and Version 2 was deployed among 41 users in Cohort 4. Table \ref{tab:data-with-avg} summarizes the number of data points for each cohort for each data type. The number of data points in Cohort 4 is greater than in Cohort 3 due to having more users in Cohort 4. On average, users have 1-3 places they go to regularly per week. In cohort 3, each user had on average 15 trips for the whole study which almost tripled in cohort 4. Also, in cohort 4 we collected about 15 feedbacks per user. Interestingly, this is much higher than the minimum required number (10) to get full compensation.

\begin{table}
    \begin{minipage}{0.45\textwidth}
    \centering
    \caption{Cohort 3 and 4 Statistics}
    \begin{tabular}{|l|c|c|c|c|}
    \hline
     Category & \multicolumn{2}{c}{\textbf{Cohort 3}} & \multicolumn{2}{c|}{\textbf{Cohort 4}} \\
     \cline{2-5}
    {} & Total & Avg & Total & Avg \\
    \hline
    Schedules    & 93   & 3.61  & 156  & 3.80 \\
    Trips        & 344  & 14.95  & 1735  & 42.30 \\
    Feedbacks    & 5    & 0.38  & 622  & 15.17 \\
    \hline
    \end{tabular}
    
    \label{tab:data-with-avg}
    \end{minipage}
\end{table}


\subsection{Scheduling Behavior Analysis}
\begin{figure}[b]
  \centering
  \includegraphics[width=0.5\textwidth]{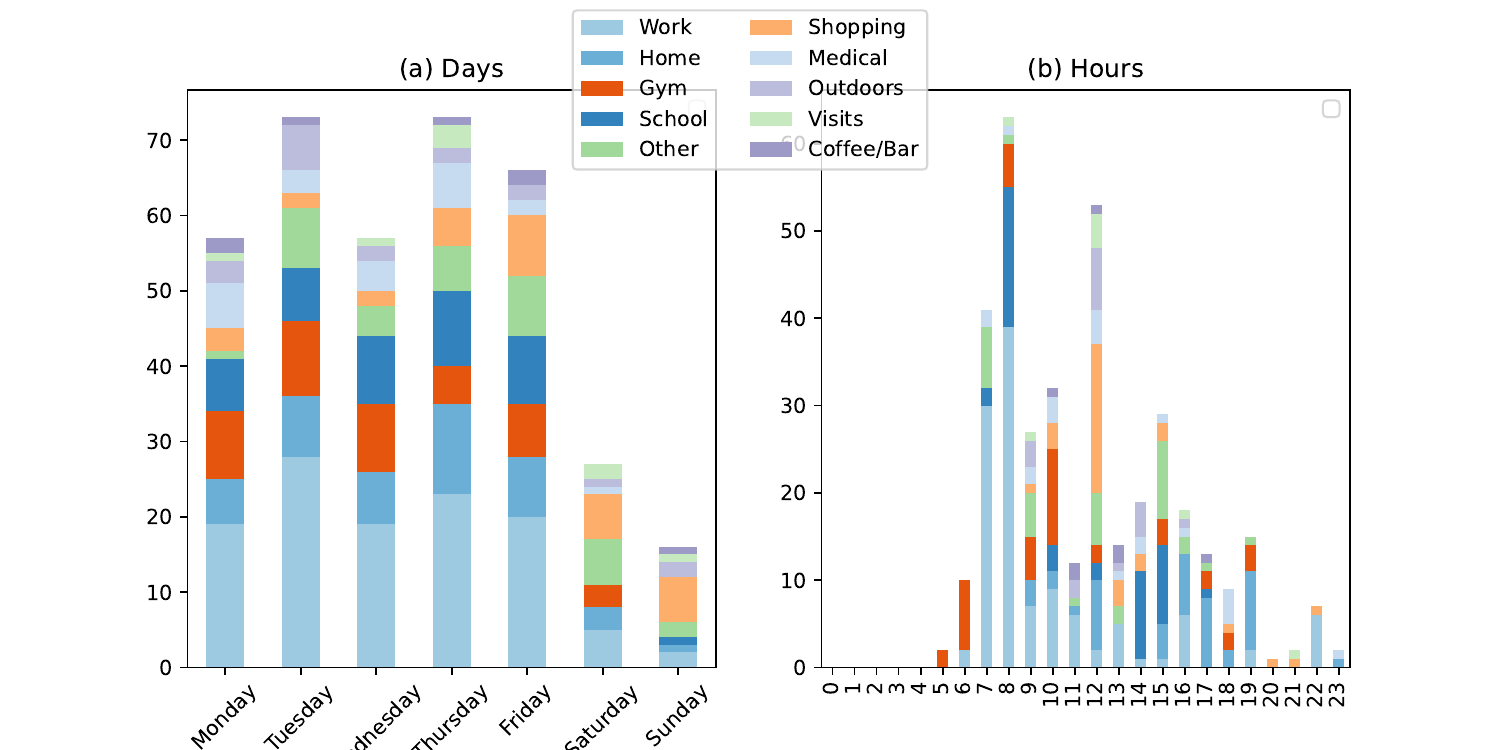}
  \caption{Most commonly scheduled places per day of the week and during the time of day.}
  \label{fig:places_combined}
\end{figure}

\begin{figure}[b]
  \centering
  \includegraphics[width=0.5\textwidth]{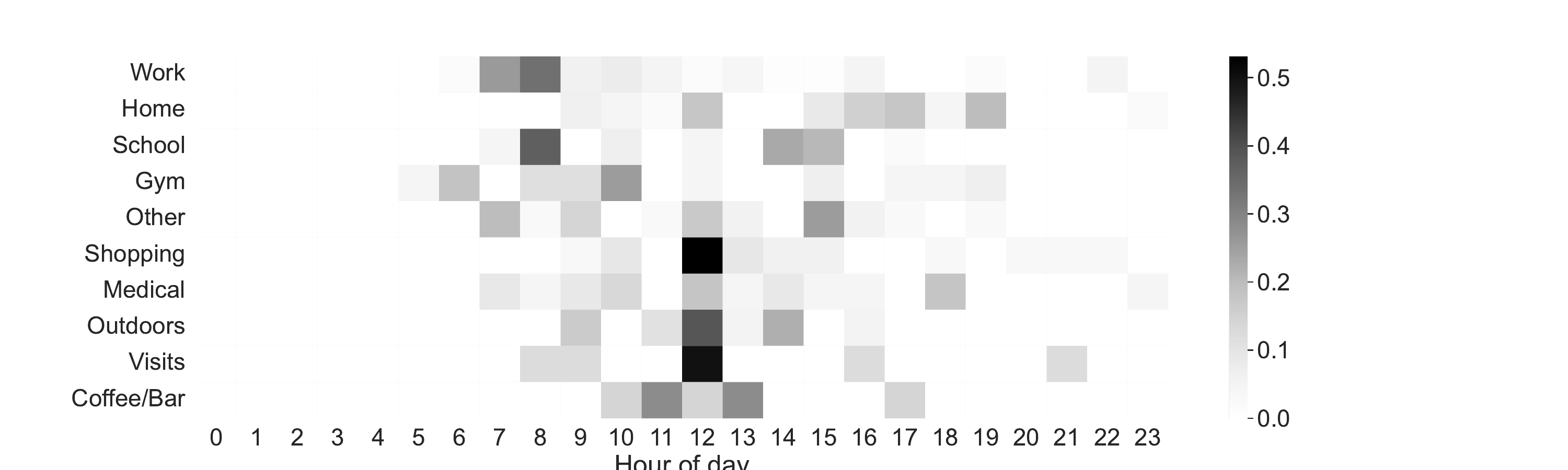}
  \caption{Places scheduling time during the day. The darker the color the more people schedule during the time.}
  \label{fig:places_heatmap}
\end{figure}

In both cohorts, people scheduled trips to a wide variety of places. The two most common places are home and work. Other places include Shopping, Gym, Medical, School, Visits, Outdoors, Coffee/Bar and others. Cohort 4 has more places, possibly because of the more schedules. Fig. \ref{fig:places_combined} (a) shows weekdays have more schedules than weekends in general, and users schedule for work and school more on weekdays and for shopping more at the end of the week. Interestingly, the Gym is more on weekdays than on weekends. Also, we see that most scheduling happens in the morning hours and the earliest place people might go to is the Gym (Fig. \ref{fig:places_combined} (b)). We can also see that shopping happens mostly at noon and that 8 am and 12 pm are the busiest times during the day. Fig. \ref{fig:places_heatmap} shows how some places are scheduled at different times during the day (lighter color) like Gym, Medical Appointment, and Home, while other places are centered around some hours (darker color) like work and school in the morning, shopping and outdoors at noon. 


\subsection{Trip Behavior Analysis}
\begin{figure}[h]
    \centering
    \includegraphics[width=0.4\textwidth]{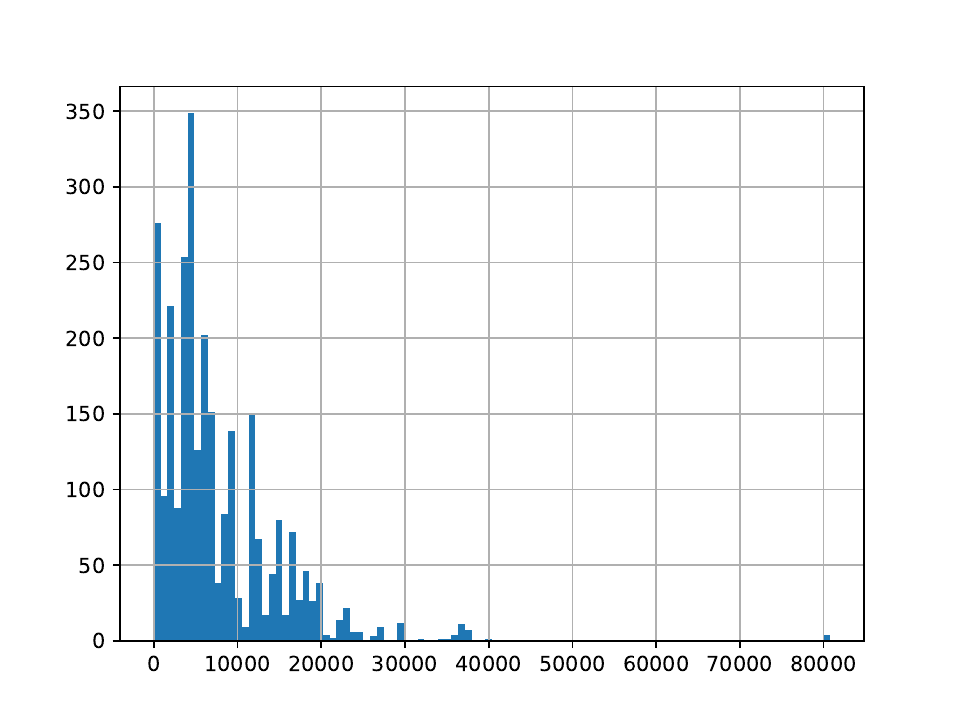}
    \caption{Distribution of the distances of all scheduled trips from the source in meters}
    \label{fig:dist_src_dst}
\end{figure}

Most trip destinations are in Denver city and few are in other cities like Boulder, Golden, and in the mountains. 
Fig. \ref{fig:dist_src_dst} shows the distribution of the distances from the source to the destination. The average distance from source to destination is 7.7 kilometers (km) and we can see that most of the locations are within 20 km. Finally, in terms of distance traveled, we found that visits had the highest average distance of 17 km while school and outdoors had the lowest (4.5 km). Other places were around the total average of 7.5 km. The difference between different groups is significant (1.92e-33).

\subsection{Feedback Analysis}

\par Looking at the feedback data, each response gave us the following information

\begin{itemize}
    \item Trip taken [Yes/No]
    \item Usefulness of the following provided information for a specific trip ( Construction, Incidents, Road Conditions, Weather, Air Quality)[1-5]
    \item Usefulness of sending a trip reminder [1-5]
    \item Which of the following information influenced their trip [1-5]
\end{itemize}

We received 622 feedbacks from 37 users. To reduce the number of random responses we filtered out responses from users who gave the same usefulness rating almost always for all types of questions. We approached that by 1) calculating the standard deviation among each response (6 questions), then 2) calculating the standard deviation again for the results on a user level. From that about 100 responses (15\%) from 3 users were excluded where the score was $<$ 0.05. 

\par We started to look at the correlation between different types of questions. Specifically, how does responding to one question affect the response to another? Overall we found that there is a slight correlation between all types of answers. For example, if someone gives a certain rating for one piece of information, they will probably give a good rating for other questions. Moreover, a stronger correlation among a couple of groups was found: (Construction, Incidents, Road conditions) and (Weather and Air Quality). For instance, if people say that the air quality information is helpful, likely, they will also say that the weather information is also helpful. Such insights can help in the design of similar questionnaires in the future.

\par
\begin{figure}[b]
  \centering
  \includegraphics[width=0.5\textwidth]{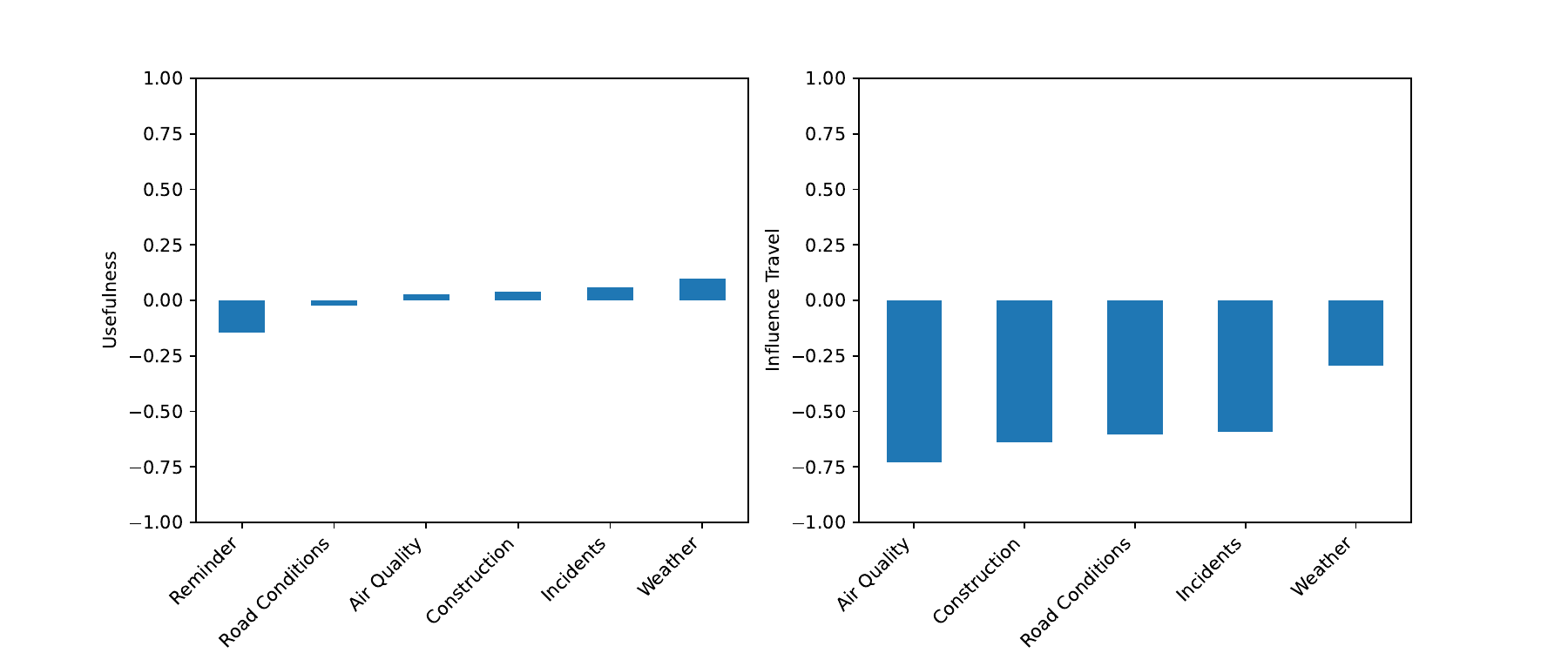}
  \caption{Average scores of usefulness and influence questions}
  \label{fig:usefulness_influence_bars}
\end{figure}

 Looking at each type of information individually, we found that on average, the perceived usefulness of most information was neutral (i.e. some people found them useful and some did not) \ref{fig:usefulness_influence_bars}. However, the difference in the perceived usefulness of each type of information was significant such that Weather information was the most useful and reminders were the least useful (ANOVA test pvalue=4.8e-05).
\par On the other hand, the average level of influence for each piece of information on the travelers' trip decision was less than the level of usefulness. However, there was a significant difference in the average level of influence among different groups (ANOVA test pvalue=7.6e-18). The least influential piece of information was air quality and the most influential was the Weather. This analysis shows that in such applications with similar scenarios, people care about certain information more than others.

\subsection{Usefulness and Influence}
\par
\begin{figure}[h]
  \centering
  \includegraphics[width=0.5\textwidth]{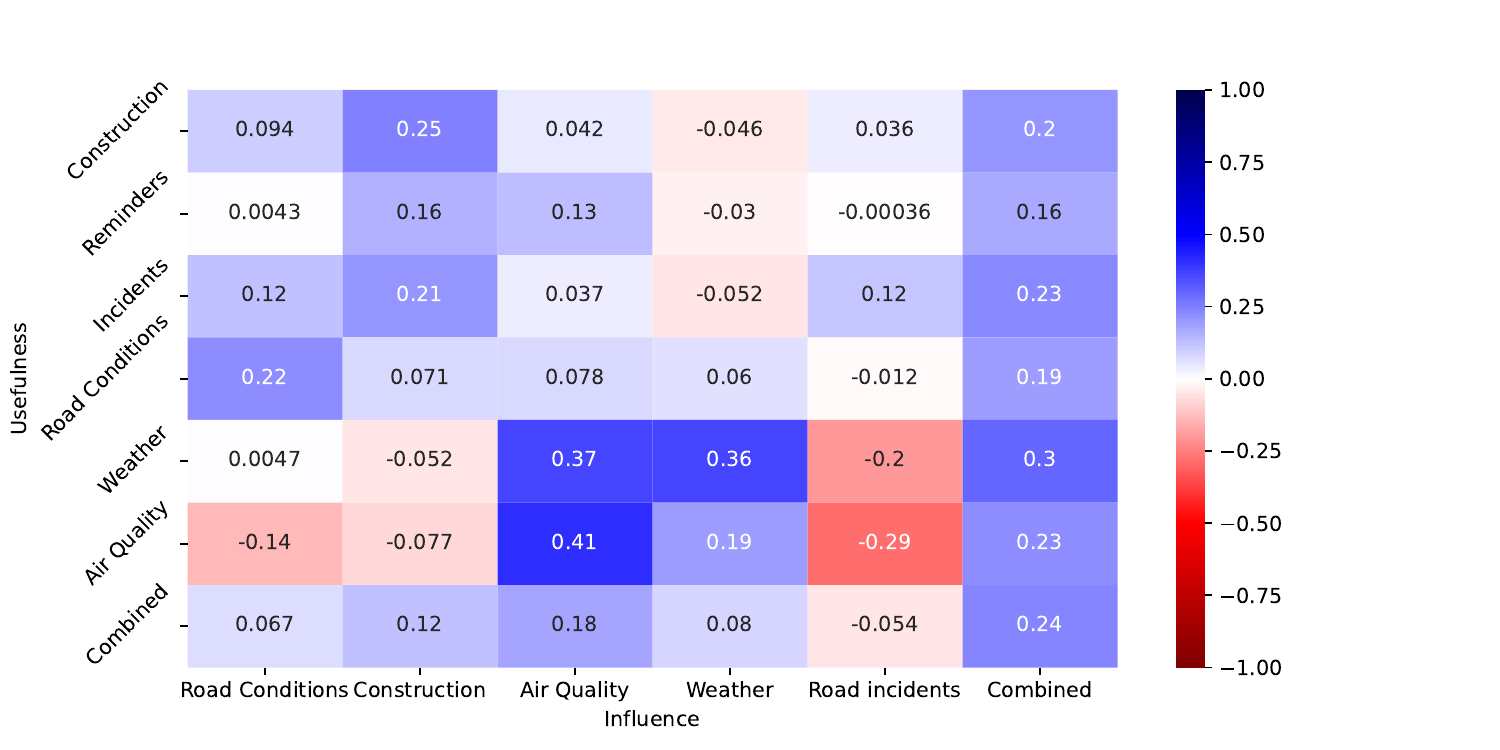}
  \caption{Correlation between usefulness and influence on trip decision}
  \label{fig:uf_inf_corr}
\end{figure}

 We conducted a correlational analysis of usefulness and influence answers and found that there is some correlation but not very strong (Fig. \ref{fig:uf_inf_corr}). If we see the rightmost column, we can see that the combined influence (whether a person was influenced by any provided information) is correlated with almost all provided information with the highest for the weather. However, the most correlated piece of information on a type level is Air quality. This says that when people say that they find one piece of information useful, it is likely that they will say that it influenced their travel decision. 

\par
\begin{figure*}
  \centering
  \includegraphics[width=\textwidth]{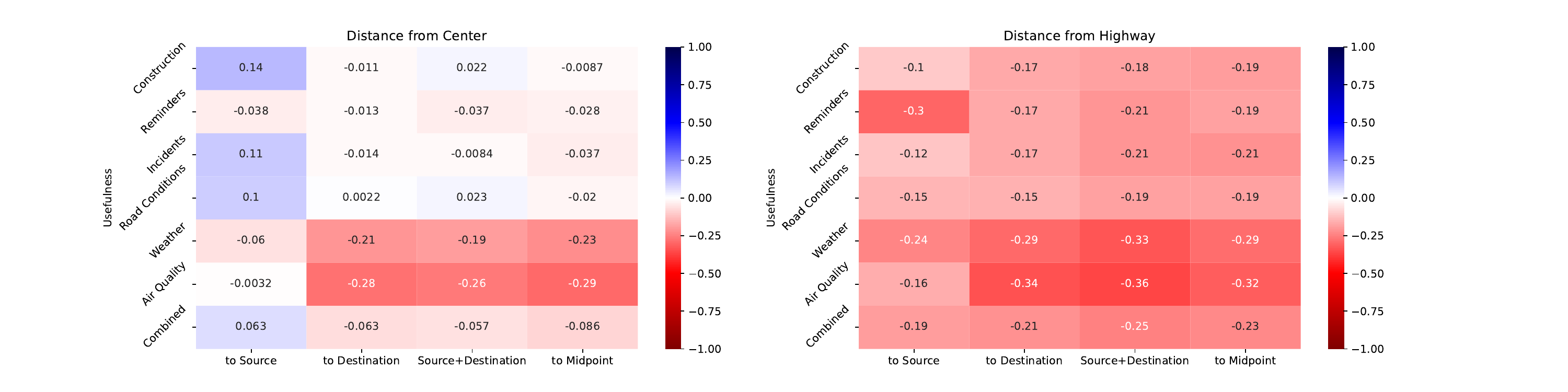}
  \caption{Correlation between usefulness and distance from the construction}
  \label{fig:distance_correlation_usefulness}
\end{figure*}

 We also asked whether distance from the construction project is a factor for people to benefit from the provided information or being influenced or not. To do that we have tested eight methods of distance as summarized in Fig. \ref{fig:distance_correlation_usefulness}. In short, we calculated the distance from 2 points (center of construction and highway) to 4 points (trip source, trip destination, trip source + trip destination, and midpoint between source and destination). Fig. \ref{fig:distance_correlation_usefulness} shows the correlation between the distance and the level of usefulness. 
\par The figure shows clearly that the closer people are to the highway, the more useful the overall information is, especially Air quality and weather information (p-value of Kendall correlation of combined usefulness score with total distance = 1.493e-09).

\subsection{Intervention-Feelings correlations}

\par The participants of PureNav also used another app that asks them daily questions about their feelings. It asks them to rate the level of the following feelings from 1 to 5: Happy, Distressed, Irritable, Alert/Awake, Lonely. To understand if PureNav usage had any correlation with a user's feelings, we conducted a correlation analysis between several feedbacks submitted by a user and their feelings (See Table \ref{tab:fb_corr}). We also averaged the number of submitted feedbacks for each feelings level (See Fig. \ref{fig:fb_avg}). Both figures show that higher usage of PureNav is positively correlated with happiness and alert/awake, and negatively correlated with Distress and Irritable. (p-values) for differences were less than 0.0001, except for loneliness. These results show that there was a relationship between the active users of Purenav and their feelings. However, The insights do not convey a causation relationship between the usage of PureNav and people's feelings. Future analysis can investigate the causation relationship between using a personalized navigation system for EJ communities and their feelings.

\begin{table}[b]
\centering
\caption{Correlation between feelings and feedbacks}
\begin{tabular}{|c|c|}
\hline
\textbf{Feeling} & \textbf{Correlation with \#feedbacks} \\
Happy & 0.164 \\
Distressed & -0.224 \\
Irritable & -0.180 \\
Alert/Awake & 0.308 \\
Lonely & -0.006 \\
\hline
\end{tabular}
\label{tab:fb_corr}
\end{table}

\section{Intervention Questionnaire Survey} 

We administered a survey at the end of each cohort to learn about user experience in using the app.
The survey questions are (Question 2 was not included in Cohort 3):

\begin{enumerate}
    \item How helpful do you find PureNav? (1 - 10) 
    \item How easy do you find PureNav to use? (1 - 10)
    \item What did you like about PureNav? (open-ended)
    \item What did you dislike about PureNav? (Open-ended)
    \item Will you continue using the app after the study period? (Yes, No)
    \item How likely are you to recommend PureNav? (1 - 10)
    \item Do you have any suggestions or comments for future releases of PureNav?
\end{enumerate}



In total, we have received 64 responses (23 from Cohort 3 and 41 from Cohort 4). 
Overall, the app's helpfulness responses are mostly negative in Cohort 3. Reasons given included unclear differences with other apps, technical issues, and destination routes being well-known. Helfulness improved significantly in Cohort 4 (from -0.21 to 0.12, 41\%) indicating that users appreciate incorporating personalized information in the app. Comments for Cohort 4 were generally positive. For easeness, the feedback received from Cohort 4 was mixed. Although there were more positive comments, some technical issues were mentioned, and the app's irrelevance for people working from home or hourly workers.




\par We asked the following questions to know how valuable the system is compared to other navigation apps:

\begin{enumerate}
    \item For local navigation within your community, what other navigation apps have you used before? (Google Maps, Waze, Apple Maps, Other)
    \item Concerning \textbf{providing useful road information} (weather, air quality, road incidents, road closures, .. etc) How valuable is PureNav compared to other apps you have used before? (1 - 10)
    \item Concerning \textbf{reminding you to leave on time} for trips that you regularly take, how valuable is PureNav compared to other apps you have used before? (1 - 10)
    \item Concerning \textbf{avoiding polluted areas}, how valuable is PureNav compared to other navigation apps? (1 - 10)
\end{enumerate}

\par Overall people found PureNav valuable in providing useful information about the construction project (Average 0.6 on a -1 to +1 scale) and helping them avoid pollution (Average 0.6).
However, they did not find reminders useful (Average -0.2). 



\section{Discussion}

\subsection{Building a helpful navigation information system}

\par Our first research question is: How do we build a helpful software system to support environmental justice communities while navigating the areas affected by large construction projects? Overall, we learned a lot, both in terms of usage and content. In terms of usage, it is essential to consider a wide range of demographics from tech-savvy young population to older generations. In terms of content, it is essential to address the issue caused by fast-paced construction projects where new road closures, construction events, and traffic changes happen very quickly and are generally not captured in traditional navigation systems. 

\par Our first version of PureNav suggested routes and provided step-by-step directions, which participants did not find useful. Some participants said, "I don't need a GPS app to go to work". Instead, based on their feedback, our second version focused on providing personalized and area-related information such as weather, construction, incidents, and air quality, which they appreciated and found useful.  

\par The key takeaway in the context of navigation is practicality. While the participants found the information we provided useful, it did not influence their travel decisions since travel is mostly necessary. In the future, we plan to improve PureNav by providing suggestions along with personalized information. For instance, when we report that the air quality is bad, we can say don't bike or walk, or wear a hat.

\par Although we know from dealing with this community previously that there are a mixture of ages including older adults, in this study we did not have specific design decisions or evaluations targeted at them. However, our design decisions and app enhancements came from analyzing the feedback as a whole from all participants.

\subsection{Trips scheduling behavior}

Our second research question is: What can we learn about Trip scheduling behavior from the usage of an information system deployed in environmental justice communities? There are several important lessons learned. First, people tend to have some (1-3) regular places that they visit often, such as Work, School, Gym, etc, and these places are generally close to their homes. In addition, scheduling is dependent on the type of place, e.g. scheduling for work and school on weekdays at specific times, while for shopping and medical appointments on weekends at varied times. Finally, people working from home or having part-time jobs have different preferences in terms of scheduling. In the future, we plan to use this knowledge to improve PureNav by optimizing personalization based on frequently visited places, work types, days of the week, and times of the day. For instance, for a person who goes to the gym at 5 PM on Fridays, PureNav can provide fine-grained information about the area around the gym or the road to the gym by accessing CCTV cameras in the vicinity a little before 5 PM on Fridays.

\subsection{Mitigating the negative impacts of the construction project}

\par Our final question is: To what extent do the proposed intervention mechanisms help in mitigating the negative impacts of the construction project for environmental justice communities who live in the affected areas? We conducted multiple investigations to answer this research question, including a correlational analysis of users' feelings and the level of usage of the system, a correlational analysis of the distance of the person to the construction and their perceived usefulness of the data and finally conducting surveys at the end of the study.

\par Our analysis shows a clear positive correlation between the level of usage of PureNav and people's level of positive feelings. Further, there is a clear negative correlation between the level of usage of PureNav and people's level of negative feelings. Next, our analysis shows a clear correlation between the closer a user is to the construction site and their perceived helpfulness of the information PureNav provides. 

\par Finally, our survey findings show that the users are generally positive towards PureNav. In particular, people were asked to rate three features that were introduced in PureNav to differentiate it from the general navigation apps. Two of them were found valuable compared to other apps which are: Providing useful road information (such as weather, air quality, road incidents, road closures, etc) and helping them to avoid polluted areas.

\par Overall, this shows that PureNav is {\it potentially} useful in improving people's well-being. It is important to note that all our analysis is based on correlation and does not establish a causal link between PureNav usage and people's well-being. Further detailed investigation is needed to establish causality.

\bibliographystyle{IEEEtran}
\bstctlcite{MyBSTcontrol}
\bibliography{references2}

\end{document}